\documentclass[10pt,jorunal]{IEEEtran} 
\usepackage[utf8]{inputenc}
\usepackage{fst}

\title{Shaped On-Off Keying Using Polar Codes}
 \author{Thomas Wiegart,~\IEEEmembership{Student~Member,~IEEE}, Fabian Steiner,~\IEEEmembership{Student~Member,~IEEE}, \\Patrick Schulte,~\IEEEmembership{Student~Member,~IEEE}, Peihong Yuan,~\IEEEmembership{Student~Member,~IEEE} 
 \thanks{T.~Wiegart, F.~Steiner, P.~Schulte, and P.~Yuan are with the Institute for Communications Engineering, Technical University of Munich (TUM). E-Mails: \texttt{\{thomas.wiegart, fabian.steiner, patrick.schulte, peihong.yuan\}@tum.de}.}\vspace*{-0.35cm}
}

\newcommand{\n}{N}

\newcommand{\UUY}{U_i|\bm{U}_1^{i-1},\bm{Y}}
\newcommand{\UU}{U_i|\bm{U}_1^{i-1}}

\newcommand{\reviewer}[1]{#1}

\newcommand{\rrreviewer}[1]{#1}
\newcommand{\fst}[1]{#1}

\providecommand\given{} 
\newcommand\SetSymbol[1][]{
   \nonscript\,#1\vert \allowbreak \nonscript\,\mathopen{}}
\DeclarePairedDelimiterX\Set[1]{\lbrace}{\rbrace}%
 { \renewcommand\given{\SetSymbol[\delimsize]} #1 }
 
 \newcommand{\pltref}[2]{\tikzset{external/export next=false}\tikz[baseline=-0.55ex]{\draw[thick, #1] (0,0) -- (0.6,0) plot[thick,mark options={solid},mark=#2] coordinates{(0.3,0)};}} 
\usetikzlibrary{external}
\tikzset{external/system call={latex \tikzexternalcheckshellescape -halt-on-error 
-interaction=batchmode -jobname "\image" "\texsource" &&
dvips -o "\image".ps "\image".dvi &&
ps2eps -f "\image.ps"}}
\begin{document}

\maketitle

\begin{abstract}
The probabilistic shaping scheme from Honda and Yamamoto \rrreviewer{(2013)} for polar codes
is used to enable power-efficient signaling for on-off keying (OOK). As OOK has a non-symmetric optimal input distribution,
shaping approaches \rrreviewer{that are based on the concatenation of a distribution matcher followed by systematic encoding do not result in optimal signaling. Instead, these approaches
represent a time sharing scheme where only a fraction of the codeword symbols is shaped.} The proposed scheme uses a polar code for joint distribution matching and forward error correction \fst{which enables asymptotically optimal signaling}. Numerical simulations show a gain of $\SI{1.8}{dB}$ compared to uniform transmission at a spectral efficiency of $\SI{0.25}{bits/channel~use}$ for a blocklength of $\SI{65536}{bits}$.
\end{abstract}


\section{Introduction}
Power efficient signaling requires a non-uniform input distribution for many channels. Combining the optimal input distribution with \ac{FEC} is not straightforward: conventional schemes (e.g., \cite[Sec.~6.2]{gallager1968}, \cite{forney_trellis_1992}) place the shaping operation after \ac{FEC} encoding so that it needs to be reversed before (or performed jointly with) the \ac{FEC} decoding. This is prone to error propagation and synchronization issues~\cite{forney_efficient_1984}.

\rrreviewer{In~\cite{ratzer2013error}, the authors build on the \emph{reverse concatenation} principle~\cite{w._g._bliss_circuitry_1981} (the shaping operation is performed \emph{before} the \ac{FEC} encoding) and introduce the concept of \emph{sparse-dense transmission}. The term ``sparse-dense'' reflects the composition of a \ac{FEC} codeword with a sparse (ones and zeros are not equally distributed) and dense part (zeros and ones are approximately uniformly distributed). The sparse part is realized with mapping techniques (e.g., look-up tables) and its distribution is maintained by systematic encoding.}

\rrreviewer{In general, any communication scheme using this approach operates in a \ac{TS} fashion as only a fraction of the codeword symbols is shaped. The explicit integration of a variable-to-fixed length \ac{DM} in a sparse-dense setup is done for the first time in \cite[Sec.~7.3]{boecherer_thesis}. The suboptimality of \ac{TS} can be circumvented by the approach of \cite[Sec.~7.4]{boecherer_thesis} which uses a chaining construction to concatenate subsequent \ac{FEC} frames. However, this is of limited practical use because of error propagation and increased latency. In \cite{git_protograph-based_2019}, the authors use sparse-dense transmission with a fixed-to-fixed length \ac{CCDM} and \ac{LDPC} codes for power efficient signaling with \ac{OOK}. Herein, gains of about \SI{1}{dB} are observed for transmission at a spectral efficiency of $\SI{0.25}{bits/channel~use~(bpcu)}$.}

Recently, \ac{PAS} was proposed~\cite{bocherer_bandwidth_2015}, which exploits the symmetry of the optimal input distribution for the \ac{AWGN} channel with a bipolar modulation format (e.g.,~ASK) such that the suboptimality of a sparse-dense scheme can be circumvented. For sign-symmetric input distributions, e.g., Gaussian or Gaussian like distributions, \ac{PAS} factors the input distribution into amplitude and sign parts that are stochastically independent. Using systematic encoding, the non-uniform distribution on the amplitudes is preserved, while the parity bits are mapped to the sign. In \cite{boecherer2019parityshaping}, \emph{syndrome shaping} is introduced, an architecture which extends PAS to arbitrary input distributions and codes with systematic encoding. However, current implementations support matching rates close to one only.

Non-coherent modulation schemes such as \ac{OOK} generally do not have a symmetric input distribution such that \ac{PAS} can not be used and schemes like \cite{git_protograph-based_2019} still exhibit a gap to capacity. In this work, we analyze a \ac{PS} approach for OOK that uses a method by Honda and Yamamoto \cite{honda2013polar,mondelli2018capacity} where polar codes \cite{St02,Ar09} perform joint distribution matching and FEC.
This idea was also applied in \cite{iscan2018shaped} with the intention to avoid an additional \ac{DM}~\cite{schulte_constant_2016} and to use a single component for distribution matching and \ac{FEC}.  
We apply this principle to \ac{OOK} and show gains of \SI{1.8}{dB} over uniform signaling at a spectral efficiency of $\SI{0.25}{bpcu}$. The proposed scheme outperforms sparse-dense signaling \cite{git_protograph-based_2019} with \ac{CCDM}.

\section{Preliminaries}
\subsection{Notation}
    Random variables are denoted by uppercase letters, e.g., $X$, while realizations or deterministic variables are denoted by lowercase letters, e.g., $x$. Vectors are denoted by a bold font, e.g., $\bm{x}$ for deterministic vectors and $\bm{X}$ for random vectors. Bold capital letters are also used for deterministic matrices. We write $\bm{x}_i^j = [x_i, \dots, x_j]$. \rrreviewer{The notation $\iH{X}$ denotes the entropy of the random variable $X$ in bits. Similarly, $\iH{X|Y}$ is the conditional entropy of  $X$ given $Y$. The \ac{MI} of $X$ and $Y$ is denoted by $\iI{X;Y}$.}
\subsection{System Model}
\begin{figure}[t]
    \centering
    \footnotesize
    \includegraphics{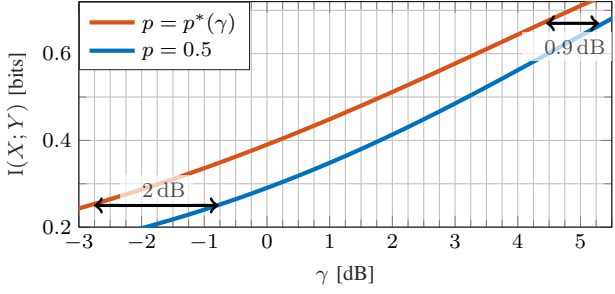}
    \caption{Achievable rates for OOK with uniform and optimized input distributions.}
    \label{fig:ook_rates}
\end{figure}
Consider an \ac{AWGN} channel
\begin{equation}
    Y = a X + N
\end{equation}
where $X \in \{0,1\}$, $a$, $N$, and $Y$ denote the transmit signal, symbol amplitude, additive white Gaussian noise, and received signal respectively. With \ac{OOK} modulation, $X$ is distributed according to
\begin{equation} \label{eq:probs}
    \pP[X][1] = p, \quad \pP[X][0] = 1 - p .
\end{equation}
The additive noise is assumed to have zero mean and unit variance. The \ac{SNR} is $\gamma = p a^2$ and an achievable rate is $\iI{X; Y}$. Fig.~\ref{fig:ook_rates} depicts $\iI{X; Y}$ versus the \ac{SNR} for two different choices of $\pP[X]$: the blue curve is for uniform $\pP[X]$ (i.e., $p = 0.5$)  and the red curve is for a $\pP[X]$ that is optimized for each SNR, i.e.,
\begin{equation}
    p^{*}(\gamma) = \argmax_{p} \quad\iI{X;Y} \quad \text{s.t.} \quad p a^2  = \gamma .
\end{equation}
There is a significant gain in power efficiency for non-uniform input symbols, e.g., for a rate of $\SI{0.25}{bpcu}$ the optimal input distribution gains approximately $\SI{2}{dB}$ over uniform inputs.

\subsection{Polar Codes} \label{sec:polar}
Polar codes \cite{St02, Ar09} are linear block codes with block length $N = 2^n$ for $n \in \mathbb{N}$ and dimension $K$. The codeword $\bm{x} \in \mathbb{F}_2^{N}$ is generated from the input $\bm{u} \in \mathbb{F}_2^{N}$ by using
\begin{align} \label{eq:polarenc}
    \bm{x} = \bm{u G}_n, \quad \text{with }   \bm{G}_n = \bm{G}_2^{\otimes n} \text{ and } \bm{G}_2 = \begin{bmatrix} 1 & 0 \\ 1 & 1\end{bmatrix}.
\end{align}
$\bm{G}_2^{\otimes n}$ denotes the $n$-th Kronecker power of $\bm{G}_2$. The codeword is transmitted over a memoryless channel $\pP[Y|X]$. The received signals are collected in the vector $\bm{y} \in \mathbb{R}^N$. The bits of $\bm{u}$ \reviewer{asymptotically} polarize into two sets \cite{Ar09}:
\begin{align}
    \cI &= \left\{ i \in \{1,\dotsc,N\} \mid \iH{\UUY} \leq \delta \right\} \label{eq:cI} \\
    \cF &= \left\{ i \in \{1,\dotsc,N\} \mid \iH{\UUY} \geq 1 - \delta \right\}
\end{align}
for $\delta > 0$. \reviewer{For finite $N$ we have a vanishing fraction of bits with $\delta < \iH{\UUY} < 1 - \delta$. }
With \ac{SC} decoding, bit $i$ in $\bm{u}$ is reliable if $i \in \cI$. 
\reviewer{Otherwise,} the bit $i$ is unreliable. The unreliable bits are \emph{frozen}, i.e., they are set to a fixed value that is known both at the encoder and the decoder. The reliable bits are used for information transmission.

Ar\i{}kan \cite{Ar09} showed that for a \ac{B-DMC}, we asymptotically have
\begin{align}
    \lim_{N \to \infty} \,\, \frac{1}{N} \left\vert \cI \right\vert &= 1 - \iH{X|Y} \\
    \lim_{N \to \infty} \,\, \frac{1}{N} \left\vert \cF \right\vert &= \iH{X|Y} .
\end{align}
\rrreviewer{For symmetric channels (i.e., $\iH{Y|X} = \iH{Y|X=x}, \forall x$), the capacity achieving distribution $P_X$ is uniform}, and we have
\begin{equation}
    \iI{X;Y} = \iH{X} - \iH{X|Y} = 1 - \iH{X|Y}.
\end{equation}
Thus polar codes achieve the symmetric capacity of \ac{B-DMC}s.

\section{Polar Codes with Non-Uniformly Distributed Codewords}
\subsection{\rrreviewer{Polarization for Non-Uniformly Distributed Codewords}}

\rrreviewer{Suppose that we want to create a codeword $\bm{x}$, where the codeword symbols have a non-uniform distribution.}
\rrreviewer{Honda and Yamamoto \cite{honda2013polar} showed that this is possible using a non-linear coding scheme based on polar codes. With the constraint on the distribution of the codewords,}
the bit positions in $\bm{u}$ do not only polarize \reviewer{asymptotically} into $\cI$ and $\cF$, but also into
\begin{align}
    \cU &= \left\{ i \in \{1,\dotsc,N\} \mid \iH{\UU} \geq 1 - \delta \right\} \label{eq:cU} \\
    \cD &= \left\{ i \in \{1,\dotsc,N\} \mid \iH{\UU} \leq \delta \right\} .
\end{align}
The $i$-th bit position of $\bm{u}$ can be used for uniform data if $i \in \cU$. If, however, $i \in \cD$, then the value of $u_i$ is (almost) deterministic given the previous values $\bm{u}_1^{i-1}$. Thus the bit positions $i \in \cD$ can not be used for data transmission, but are frozen to a value that depends (non-linearly) on the previous input. We describe the encoding procedure in Sec.~\ref{sec:encoding}.

\begin{figure}[t]
    \centering
    \includegraphics{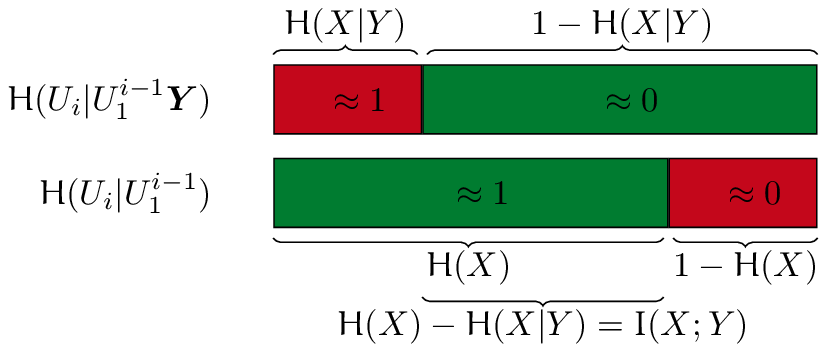}
    \caption{Graphical representation of the polarization of $\iH{\UUY}$ and $\iH{\UU}$ in $\bm{u}$ (ordered).}
    \label{fig:PC_U}
\end{figure}

In \cite{honda2013polar}, it was additionally shown that
\begin{align}
    \lim_{N \to \infty} \,\, \frac{1}{N} \left\vert \cU \right\vert &= \iH{X}, \\
    \lim_{N \to \infty} \,\, \frac{1}{N} \left\vert \cD \right\vert &= 1 - \iH{X}.
\end{align}
Therefore, the fraction of bits that can be used for uniform data is asymptotically $\iH{X}$.
Fig.~\ref{fig:PC_U} shows a graphical representation of the input $\bm{U}$. The fraction of bits that can be transmitted reliably (i.e., where $\iH{\UUY} \approx 0$) is (asymptotically) $1 - \iH{X|Y}$ and the fraction of bits that can be used for uniform data such that a shaped codeword can be obtained is (asymptotically) $\iH{X}$. Conditioning does not increase entropy and thus $\iH{\UUY} \leq \iH{\UU}$. It follows that for a bit at position $i$ with $\iH{\UUY} \approx 1$, we also have $\iH{\UU} \approx 1$, i.e., \rrreviewer{$i \in \cF$ implies $i \in \cU$ and thus $\cF \subset \cU$.}

\rrreviewer{A bit $i$ in $\bm{u}$ can be used for information if it is reliable (i.e., $i \in \cI$) and if uniform data is allowed at this position (i.e., $i \in \cU$).} The set of bits that can be used for information transmission is thus $\cU \cap \cI = \cU \setminus \cF$ with
\begin{equation}
    \lim_{N \to \infty} \,\, \frac{1}{N} \left\vert \cU \cap \cI \right\vert = \iH{X} - \iH{X|Y} = \iI{X;Y}
\end{equation}
and the scheme can achieve capacity on asymmetric \ac{B-DMC}s~\cite{honda2013polar}.

\rrreviewer{In the procedure of encoding, decoding, and code construction we will handle three different types of bit positions in the input $\bm{u}$:
\begin{itemize}
    \item If $i \in \cU \cap \cI$, bit position $i$ will be used for uniform data.
    \item If $i \in \cF$, bit position $i$ will be frozen to a value known to the encoder and the decoder.
    \item If $i \in \cD$, bit position $i$ will be set to a value depending on the previous input $\bm{u}_1^{i-1}$ during encoding. The value is not known to the decoder.
\end{itemize}
}

\subsection{Encoding} \label{sec:encoding}
The requirement on $\pP[X][x]$ induces a constraint on the joint distribution $\pP[\bm{U}][\bm{u}]$. The task of the encoder is to generate a $\bm{u}$ that contains data and fulfills this constraint. The codeword $\bm{x}$ is generated from $\bm{u}$ as in \eqref{eq:polarenc}. 
Honda and Yamamoto \cite{honda2013polar} observed that $\pP[\bm{U}][\bm{u}]$ can be calculated efficiently using a polar decoder. Using the chain rule, $\pP[\bm{U}][\bm{u}]$ can be decomposed as
\begin{equation}
    \pP[\bm{U}][\bm{u}] = \prod_{i=1}^{\n} \pP[U_i | \bm{U}_1^{i-1}][u_i | \bm{u}_1^{i-1}] .
\end{equation}
When a \ac{SC} polar decoder is initialized just with information on the distribution of $\bm{x}$, i.e., with a \ac{LLR} $L = \log(\pP[X][0] / \pP[X][1])$, it outputs for bit position $i$ the probability $\pP[U_i | \bm{U}_1^{i-1}][u_i | \bm{u}_1^{i-1}]$ given a realization of $\bm{u}_1^{i-1}$.

Honda and Yamamoto \cite{honda2013polar} thus proposed an encoding scheme that successively encodes bit by bit as follows: \rrreviewer{If $i \in \cU \cap \cI$, then $u_i$ is used for (uniform) data. If $i \in \cF$, $u_i$ is chosen from a uniform distribution and the value is assumed to be known at the decoder as well (the value can be chosen once and kept constant for every block). Otherwise (i.e., if $i \in \cD$), $u_i$ is set according to}
\begin{equation} \label{eq:randomized}
    u_i = \begin{cases}
        0 & \text{with probability } \pP[U_i | \bm{U}_1^{i-1}][0|\bm{u}_1^{i-1}], \\
        1 & \text{with probability } \pP[U_i | \bm{U}_1^{i-1}][1|\bm{u}_1^{i-1}].
    \end{cases}
\end{equation}
This method is called \emph{randomized rounding rule} in \cite{mondelli2018capacity}. 

A simplified approach is an encoding rule called the \emph{argmax rule} in \cite{mondelli2018capacity}. Here, \rrreviewer{for the values of $u_i$ with $i \in \cD$,} one chooses
\begin{equation} \label{eq:argmax}
    u_i = \begin{cases}
        0 & \text{if } \pP[U_i | \bm{U}_1^{i-1}][0|\bm{u}_1^{i-1}] \geq \pP[U_i | \bm{U}_1^{i-1}][1|\bm{u}_1^{i-1}], \\
        1 & \text{else}.
    \end{cases}
\end{equation}

The randomized coding rule yields provable capacity achieving results, whereas the argmax rule yields better finite length results \cite{mondelli2018capacity,chou2015deterministic} \rrreviewer{and does not need any randomness for encoding or decoding.}
\fst{\subsection{List Encoding}
\label{sec:list_encoding}
During successive encoding a hard decision for the bits $u_i$ with $i \in \cD$ must be done using \eqref{eq:randomized} or \eqref{eq:argmax}. This hard decision may not be ideal especially for bit positions where $\iH{\UU}$ is not polarized perfectly. One could thus follow the idea of \cite{TaVa15} and use a \ac{SCL} decoder for encoding that branches a list when a hard decision is done. This idea was also applied in \cite{iscan2018shaped}. The list can be pruned with the usual metric used for \ac{SCL} decoding. At the end, the \ac{SCL} encoder outputs a list of valid codewords, i.e., all codewords contain the encoded data. We choose the codeword that has an empirical distribution closest to the target distribution.}

\subsection{Decoding} \label{sec:decoding}
The decoder estimates $\bm{u}$ from the noisy channel observations $\bm{y}$. The estimates are stored in a vector $\hat{\bm{u}}$.
Decoding is performed with a \ac{SC} or \ac{SCL} decoder \cite{TaVa15}. 

\rrreviewer{To show the capacity achieving property it is assumed in \cite{honda2013polar} that the decoder has knowledge about the values of $u_i$ with $i \in \cD$. This knowledge can be obtained by running a \ac{SC} decoder initialized with the \ac{LLR} $L = \log(\pP[X][0] / \pP[X][1])$ that mimics the encoder and successively calculates the probabilities $\pP[U_i | \bm{U}_1^{i-1}][u_i|\hat{\bm{u}}_1^{i-1}]$ (it is assumed that previous bits have been decoded correctly, i.e., $\bm{\hat{u}}_1^{i-1} = \bm{u}_1^{i-1}$). The random choices of \eqref{eq:randomized} can be recovered by using a pseudo-random number generator at the encoder and the decoder that is initialized with the same seed.}

\rrreviewer{Simplifications are possible: As $\iH{\UUY} \leq \iH{\UU}$, it follows that if $\iH{\UU}$ is close to zero (i.e., if $i \in \cD$), then also $\iH{\UUY}$ is close to zero (i.e., $i \in \cI$) and the bits at position $i$ with $i \in \cD$ can be estimated reliably without running a second decoder.
Thus, a practical \ac{SC} or \ac{SCL} decoder implementation works as follows: if $i \in \cF$, then $\hat{u}_i$ is set to the known frozen value. Otherwise (i.e., if $i \in \cU \cap \cI$ or if $i \in \cD$), $\hat{u}_i$ is decoded regularly.}
This idea is also used in \cite{iscan2018shaped}, and it keeps the complexity at the receiver almost identical to a receiver for uniformly distributed codewords. 

\subsection{Code Construction}
\reviewer{Code construction consists of finding the four sets $\cI$, $\cF$, $\cU$, and $\cD$. For finite length simulations, we slightly deviate from the definitions in \eqref{eq:cI} and \eqref{eq:cU} and pursue the following strategy: we choose the sets $\cD$ and $\cF$. Then, $\cU$ and $\cI$ are given by $\cU = \{1,\dots,\n\} \setminus \cD$ and $\cI = \{1,\dots,\n\} \setminus \cF$, respectively.
To choose $\cD$ and $\cF$, we first estimate an ordering of $\iH{\UU}$ and $\iH{\UUY}$, respectively. Second, for a fixed transmission rate $R$, we find a tradeoff between the size of $\cD$ and $\cF$ such that $\vert\cD\vert + \vert\cF\vert = \n ( 1 - R)$.}

\reviewer{We use a Monte Carlo approach to estimate the ordering of $\iH{\UU}$ and $\iH{\UUY}$ as described in the following. We remark that one can extend the Tal-Vardy construction \cite{Tal2013construct} by using the method of \cite{kartowsky2017greedymerge} to estimate the entropy values with less computational effort. To estimate $\iH{\UU}$ with the Monte Carlo approach, a SC decoder is initialized with the LLR $\log(\pP[X][0] / \pP[X][1])$. When choosing the inputs $u_i$ successively using the randomized rounding rule, the \ac{SC} decoder successively outputs $\pP[U_i | \bm{U}_1^{i-1}][u_i | \bm{u}_1^{i-1}]$. Sampling over many frames, one can estimate $\iH{\UU}$. Furthermore, a transmission over the channel with the randomly generated data is simulated and a SC decoder is applied. If the decoder produces a wrong decision for bit $i$, the error counter for this bit position is increased by one and the error is corrected. After many trials, the error counter for each bit position gives a reliability order for the bit positions and --- as the entropy $\iH{\UUY}$ is a monotone function of the error rate of bit position $i$ --- an order for the entropies $\iH{\UUY}$.}

\reviewer{We now choose the $D$ bits with lowest $\iH{\UU}$ to form the set $\cD$ and the $\n (1-R) - D$ bits with highest $\iH{\UUY}$ to form the set $\cF$. The optimal $D$ can be found by numerical simulations. Numerical results show that one has to choose a $D$ that is only slightly higher than the asymptotic limit $\n (1-\iH{X})$ for good results. Depending on the choice of $D$, there is a slight mismatch between the target distribution and the empirical distribution in $\bm{x}$. This stems from the finite length rate loss of the \ac{DM} process, which is discussed in Sec.~\ref{sec:rateloss}.}

\section{Numerical Results} \label{sec:results}

\subsection{Finite Length Rate Loss Evaluation} \label{sec:rateloss}

\begin{figure}[t]
    \centering
    \footnotesize
    \includegraphics{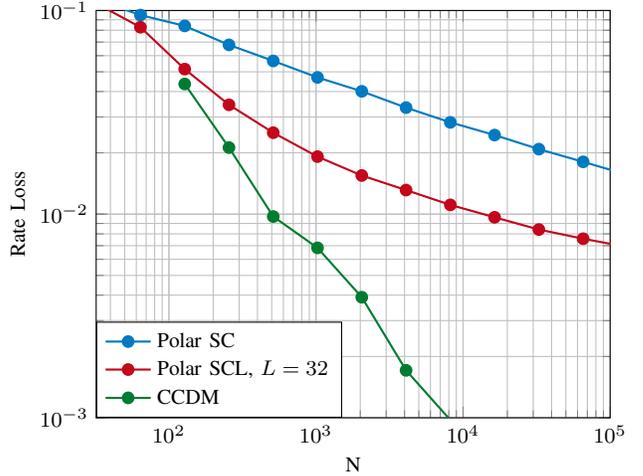}
    \caption{Rate loss comparisons for different \ac{DM} architectures. The \ac{DM}s have a binary output alphabet and a matching rate of \SI{0.5}{bits/output~symbol}.} 
    \label{fig:rateloss}
\end{figure}

\rrreviewer{The rate loss~\cite[Sec.~V-B]{bocherer_bandwidth_2015} is an important metric to analyze the finite length performance of a \ac{DM} scheme. Assume an output blocklength of $N$ bits. The rate loss is then defined as
\begin{align}
    R_\text{loss} = \iH{X} - \frac{\abs{\cU}}{N}\label{eq:rate_loss}
\end{align}
for the polar \ac{DM}. For the \ac{CCDM}, we have 
\begin{align}
    R_\text{loss} = \iH{X} - \left\lfloor\log_2\binom{N}{P_X(0)\cdot N}\right\rfloor/N\label{eq:rate_loss_ccdm}.
\end{align}
We numerically characterize $R_\text{loss}$ for different \ac{DM} architectures in Fig.~\ref{fig:rateloss}. For this, we fix a desired \ac{DM} rate of $\SI{0.5}{bits/output~symbol}$ and evaluate \eqref{eq:rate_loss} for different blocklengths $N$. We observe that \ac{CCDM} is superior to the polar \ac{DM} for all considered lengths. Its superior performance for long blocklengths is to be expected from previous results~\cite{schulte_divergence_2017}, which showed the optimality of \ac{CCDM} for fixed-to-fixed matching and $N\to\infty$. We remark that the polar \ac{DM} rate loss can be decreased if the list encoding of Sec.~\ref{sec:list_encoding} is used, see Fig.~\ref{fig:rateloss}.}

\subsection{Coded Results}

We evaluate the performance of the presented transmission scheme. 
For fixed $D$ we estimate the empirical codeword distribution and scale the amplitude $a$ so that we transmit at the target SNR, i.e., we choose $a$ such that
\begin{equation}
    \frac{1}{\n} \sum_{i=1}^{\n} (a \opE[X_i])^2 = \gamma .
\end{equation}

\begin{figure}[t]
    \centering
    \footnotesize
    \includegraphics{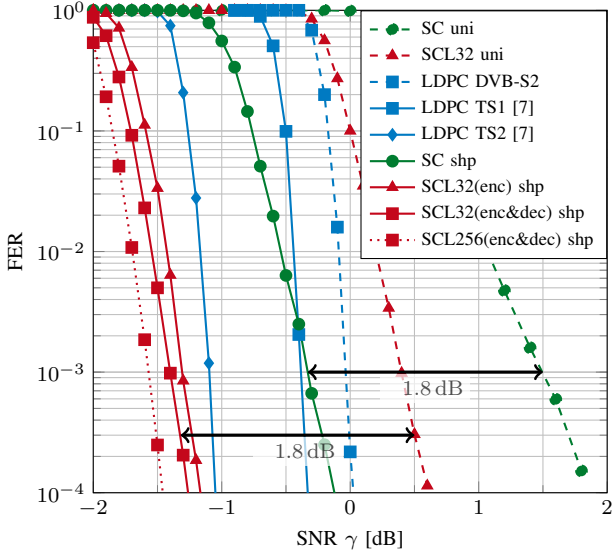}
    \caption[]{Comparison of shaped and uniform polar and LDPC codes for $\n = \SI{65536}{bits}$ and transmission rate $R = \SI{0.25}{\bpcu}$. The polar codes with \ac{SCL} decoding are combined with an outer CRC of length $\SI{32}{bits}$. They were designed at a SNR of $\SI{-1.25}{dB}$. For the curve with SC encoding and SCL decoding (list size $L=32$,
    \pltref{TUMred}{triangle*})
    we chose $D = \num{25500}$ and performed encoding using the argmax rule. For the curve with SCL encoding and decoding (both list size $L=32$, 
    \pltref{TUMred}{square*})
    we used $D = \num{25000}$. The performance can be enhanced further by increasing the list size, e.g., to $L=256$ for encoding and decoding 
    (\pltref{TUMred, dotted}{square*}).}
    \label{fig:res_65k_0.25}
\end{figure}

Fig.~\ref{fig:res_65k_0.25} shows a numerical example for $\n = \num{65536}$ and transmission rate $R = 0.25$. At this rate gains up to $\SI{2}{dB}$ can be expected from Fig.~\ref{fig:ook_rates}. With \ac{SC} decoding, the shaped polar code 
(\pltref{TUMgreen}{*})
gains about $\SI{1.8}{dB}$ at a \ac{FER} of $10^{-3}$ compared to the polar code with uniform codewords
(\pltref{TUMgreen,dashed}{*}).
With \ac{SCL} encoding and decoding (both with list size $32$) and an outer CRC, the shaped polar code
(\pltref{TUMred}{square*})
gains around $\SI{1.8}{dB}$ compared to the uniform reference
(\pltref{TUMred, dashed}{triangle*}).
The performance of the polar code at this blocklength is limited by the relatively small list size. Increasing the list size can further improve the performance, e.g., when choosing $L=256$
(\pltref{TUMred, dotted}{square*})
the performance improves by $\SI{0.2}{dB}$ compared to $L=32$.
We also include the performance of LDPC codes with blocklength \num{64800} bits using the time-sharing based \ac{PS} scheme from \cite{git_protograph-based_2019}. The LDPC code with uniform signaling
(\pltref{TUMBlue,dashed}{square*})
is taken from the DVB-S2 standard~\cite{etsi2009dvb}. 
The difference between the TS1
(\pltref{TUMBlue}{square*})
and TS2
(\pltref{TUMBlue}{diamond*})
code is that TS2 uses different signaling amplitudes on the systematic and parity parts. Both codes have been optimized individually for the respective scenario.
A \ac{CCDM}~\cite{schulte_constant_2016} is used in both cases as a \ac{DM}.

\begin{figure}[t]
    \centering
    \footnotesize
    \includegraphics{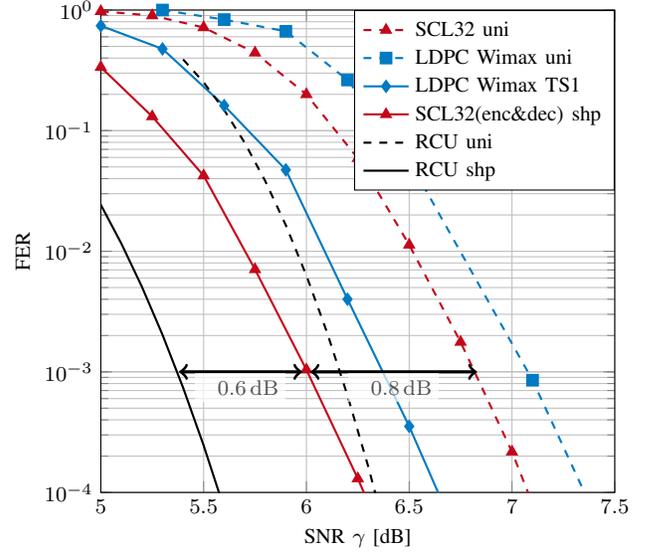}
    \caption{Comparison of shaped and uniform polar and LDPC codes for $\n = \SI{1024}{bits}$ and a transmission rate $R = \SI{0.67}{\bpcu}$. The polar codes with \ac{SCL} decoding use a list of size 32 for encoding and decoding and an outer CRC of length $\SI{16}{bits}$. They were designed at a SNR of $\SI{6}{dB}$ and we chose $D = 115$.}
    \label{fig:res_1024_0.67}
\end{figure}

In Fig.~\ref{fig:res_1024_0.67}, we depict the performance for a scenario with $R=2/3$, where gains up to $\SI{0.9}{dB}$ can be expected. The polar codes (\pltref{TUMred}{triangle*}:
 shaped,
 \pltref{TUMred,dashed}{triangle*}: uniform) have a blocklength of \SI{1024}{bits}, while the reference \ac{LDPC} codes from the Wimax standard~\cite{ieee_wimax} have a blocklength of \SI{1056}{bits} and code rates of 2/3
 (\pltref{TUMBlue, dashed}{square*}:
 uniform) and 3/4
 (\pltref{TUMBlue}{diamond*}:
 shaped). \rrreviewer{We depict a curve for TS1 only as it turns out (both by achievable rate analysis and finite length simulations) that the gain of TS2 over TS1 vanishes with increasing rate \cite{git_protograph-based_2019}}. As expected from previous works~\cite{liva_code_2016}, polar codes with SCL show an excellent performance for short to medium blocks. 
In all \ac{LDPC} cases, two hundred belief propagation iterations are performed. We also include two finite length random coding union (RCU) bounds based on saddlepoint approximations of the RCU bound \cite{font-segura_saddlepoint_2018}. At a \ac{FER} of \num{e-3}, we operate within \SI{0.6}{dB} of these bounds.

\section{Conclusion}
We applied the shaping scheme by Honda and Yamamoto for polar codes \cite{honda2013polar} to OOK transmission. Compared to previous approaches, the proposed scheme is asymptotically optimal and shows superior performance for finite length. Especially for low transmission rates, the performance is substantially better than a \ac{TS} based LDPC implementation. \rrreviewer{Future work may also compare shaped \ac{OOK} to pulse position modulation based schemes such as \cite{Donev2018PolarCodedPP} with a multilevel coding/multistage decoding architecture.}

\section*{Acknowledgements}
The authors would like to thank Ido Tal and Boaz Shuval for motivating this study, as well as Gerhard Kramer for helpful comments and discussions. 
\fst{The authors would also like to thank the anonymous reviewers who provided valuable input and ideas for improvement.}

\bibliographystyle{IEEEtran}
\bibliography{IEEEabrv,confs-jrnls,literature}

\enlargethispage{-5cm}
\end{document}